\documentclass[aps,twocolumn,letterpaper,10pt,amsmath,amssymb,prl,altaffilletter,showpacs]{revtex4-1} 
\usepackage[greek,USenglish]{babel}
\usepackage[T1]{fontenc}
\newcommand{\upmu}{\textnormal{\greektext m\latintext}} 

\usepackage{bm,mathrsfs,dcolumn,graphicx,color}
\usepackage{units,setspace}

\begin{document}
\singlespacing

\title{Magnetic control of Coulomb scattering and terahertz transitions among excitons}

\date{\today}

\pacs{78.55.Cr, 71.35.-y, 78.67.De, 75.75.-c}

\normalsize
\author{J.~Bhattacharyya,$^{1}$ S.~Zybell,$^{1,2}$ F.~E\ss er,$^{1,2}$ M.~Helm,$^{1,2}$ and H.~Schneider,$^{1,*}$ \\
L. Schneebeli,$^3$ C.~N.~B\"ottge,$^3$ B.~Breddermann,$^3$ M.~Kira,$^3$ and S.~W.~Koch,$^3$ \\
A.~M.~Andrews,$^4$ and G.~Strasser$^4$ \\
\small{\textit{$^1$Helmholtz-Zentrum Dresden-Rossendorf, P.O. Box 510119, 01314 Dresden, Germany}} \\
\small{\textit{$^2$Technische Universit\"at Dresden, 01062 Dresden, Germany}} \\
\small{\textit{$^3$Department of Physics and Material Sciences Center, Philipps-Universit\"at Marburg, Renthof 5, 35032 Marburg, Germany}} \\
\small{\textit{$^4$Institute of Solid State Electronics, Technische Universit\"at Wien, Floragasse 7, 1040 Wien, Austria}} \\
\footnotesize{$^*$corresponding author: h.schneider@hzdr.de}\\[-1ex]
}

\begin{abstract}
Time-resolved terahertz quenching studies of the magnetoexcitonic photoluminescence from GaAs/AlGaAs quantum wells 
are performed. A microscopic theory is developed to analyze the experiments. Detailed experiment--theory comparisons reveal a remarkable magnetic-field controllability of the Coulomb and terahertz interactions in the excitonic system. 
\end{abstract}

\maketitle

The exciton is a Coulomb-bound electron--hole pair that has similarity to a hydrogen atom~\cite{SQObook2011}. Even though excitons in semiconductors only have binding energies in the terahertz ($\unit[1]{THz}\:\widehat{=}\:\unit[4.1]{meV}$) 
range, they strongly influence interband optical properties~\cite{PhysRevLett.53.2173,SchmittRink.adv.phys1989,Koch2006Semiconductor}, which can critically alter the characteristics of optoelectronic and photonic devices~\cite{Warburton2000, Xia2008, Gabor2009}. Therefore, nanotechnology applications may be significantly expanded if one is able to systematically control excitons and their fundamental interactions. 

A magnetic field can considerably modify both electronic and excitonic properties. For example, free electrons become bound to Landau levels that can be detected as the quantum Hall effect~\cite{Kohl1989,Bloch2008}. Furthermore, a magnetic field enhances the binding of excitons; the resulting magnetoexcitons have a reduced Bohr radius and scattering time. This feature has been utilized, e.g., to reach the regime of superradiance~\cite{Noe2012} with the help of a magnetic field.
 
Excitons themselves can directly be accessed by terahertz (THz) spectroscopy, revealing, e.g., the presence of exciton populations, their formation dynamics, and the internal interaction properties~\cite{CernePRL1996,SalibPRL1996,KonoPRL1997}. Dressing excitons by strong THz fields leads to interesting nonperturbative phenomena like excitonic Rabi oscillations~\cite{WagnerPRL2010} and high-order sideband generation~\cite{ZaksNAT2012Recollisions}. 
In addition, light--matter coupling can be enhanced further through a resonant microcavity, transforming the polaritonic 1$s$ and the optically dark 2$p$ states into a $\Lambda$ system~\cite{Tomaino2012}. 
Furthermore, it has been shown that the Coulomb interaction effectively couples excited exciton states leading to transitions that are dipole-forbidden in a non-interacting system~\cite{Rice2013}.

In this Letter, we show that a magnetic field ($B$-field) can be used efficiently to externally control the THz-induced intra-excitonic coupling dynamics. Whereas very strong $B$-fields inhibit exciton formation because the electrons and holes are confined to quantized Landau orbitals, 
the combined effects of Coulomb- and $B$-field interactions lead to a new dressed state for weak to moderate magnetic-field strengths. In these magneto\-excitons, the energies and wave functions are modified relative to the field-free case and, in analogy to the Zeeman splitting, the degeneracy of excitons with different magnetic quantum numbers $m$ is lifted~\cite{STARK1990,HaugKochBook}.

\begin{figure}
\centering
\includegraphics[width=0.8 \columnwidth]{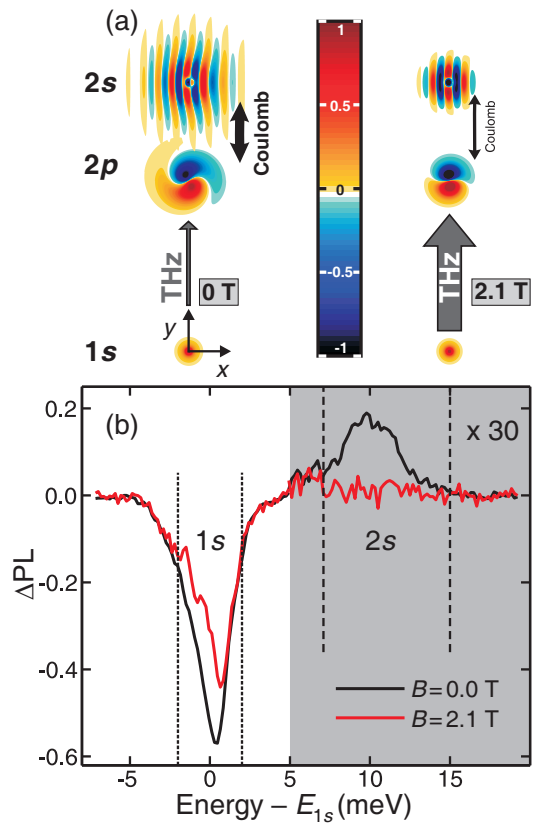}
\caption{(color online).
Influence of a magnetic field on THz- and Coulomb interactions.
(a) Computed exciton wave functions. Horizontal and vertical directions represent spatial coordinates. Note that the 2$p$ and 2$s$ energies (not shown) are $B$-dependent and presented in Fig.~\ref{Fig: 1s-np transitions and experiment}(a); for details, see text.
(b) Measured differential PL spectrum for $\hbar \omega_{\mathrm{THz}} = \unit[8.7]{meV}$. The vertical lines mark the spectral integration area around the 1$s$ and 2$s$ emission energies.
}
\label{Fig: intro}
\end{figure}

Figure~\ref{Fig: intro}(a) visualizes the interplay of $B$-field, Coulomb scattering, and THz-induced transitions among the ground state (1$s$) and two excited states (2$p$, 2$s$).
As shown in Ref.~\cite{Rice2013}, the Coulomb scattering can co-operate with THz transitions to produce an efficient transfer between 2$p$ and 2$s$ excitons when the THz field is resonant with the 1$s$-to-2$p$ transition. 
More specifically, the Coulomb interaction induces a momentum change to the 2$s$ excitons such that they are displaced with respect to the 2$p$ excitons after a Coulomb-scattering event. The rate of such scattering is determined by the spatial overlap between a stationary 2$p$ and a
displaced 2$s$ exciton wave function. We will show that this overlap, and thus the 2$p$-to-2$s$ scattering rate, can be significantly modified in the presence of a magnetic field.

To illustrate this effect, we show in Fig.~\ref{Fig: intro}(a) contours of exemplary stationary $\lambda=(1s,\mathbf{0})$ (bottom) and $\lambda=(2p,\mathbf{0})$ (middle) as well as 
displaced $\lambda=(2s,\mathbf{q})$ (top) exciton wave functions $\varphi_\lambda(\mathbf{r})$ without magnetic field (left) and with (right) a $B=\unit[2.1]{T}$ field. The Coulomb scattering induces a momentum displacement $\hbar \mathbf{q}$ to 
this 
2$s$ wave function, i.e., $\varphi_{2s,\mathbf{q}}(\mathbf{r}) = \varphi_{2s,\mathbf{0}}(\mathbf{r}) \, \mathrm{e}^{\mathrm{i} \mathbf{q}\cdot \mathbf{r}}$.
The spiral shape of the stationary 2$p$ wave function originates from density-dependent terms within the generalized Wannier equation~\cite{SQObook2011} that defines the exciton wave function and the ``interference pattern'' in the 
2$s$ scattering states 
stems from its $\mathrm{e}^{\mathrm{i} \mathbf{q}\cdot \mathbf{r}}$ part.
We have moved the 1$s$, 2$p$, and 2$s$ wave functions in the $y$ direction to enhance the visibility. For vanishing $B$-field, $\varphi_{2s,\mathbf{q}}$ and $\varphi_{2p,\mathbf{0}}$ have a large overlap integral since the plane-wave part removes their orthogonality. However, already at $B=\unit[2.1]{T}$, the magnetic-field effects dominate over Coulomb effects, as seen from the weakened interference pattern in $\varphi_{2s,\mathbf{q}}$ which is strongly contracted as well. Hence, the magnetic field tends to make 2$p$ and displaced 2$s$ states orthogonal, yielding a strongly reduced Coulombic scattering strength. At the same time, the 1$s$-to-2$p$ THz-transition strength (gray arrow) increases for elevated $B$-fields due to the contraction of the exciton wave functions.
Hence, we should be able to gradually switch off the 2$p$-to-2$s$ scattering by increasing the $B$-field.

To test this hypothesis, we perform experiments where the 1$s$-to-2$p$ transition of quantum well (QW) excitons is excited by a free-electron laser, emitting wavelength-tunable (\unit[3--200]{$\upmu$m}), picosecond~(ps)-long THz pulses. 
These excitons are produced by near-infrared (NIR) interband excitation at a fluence of $\unit[0.15]{\mathrm{\upmu J/cm^{2}}}$ using a Ti:sapphire laser emitting $\unit[4]{ps}$
pulses at $\unit[1.627]{eV}$. As sample, we use a high-quality multiple QW structure grown by molecular beam epitaxy on
a semi-insulating GaAs substrate. The sample comprises 60 GaAs QWs of $\unit[8.2]{nm}$ width separated by $\unit[19.6]{nm}$ wide AlGaAs
barriers. The heavy-hole 1$s$ excitonic state, the lowest energy level in the QWs, is at $\unit[1.566]{eV}$ with a linewidth
of $\unit[3]{meV}$, as inferred from absorption measurements at $\unit[10]{K}$, while the quasi-degenerate 2$s$ and 2$p$ excitonic states are located at $\unit[1.575]{eV}$. The light-hole excitonic 1$s$ energy is still higher, at $\unit[1.583]{eV}$~\cite{WagnerPRL2010}. We focus
both lasers onto the sample and detect the photoluminescence (PL) by a synchroscan streak camera~\cite{BhattacharyyaRSI2011}. Allowing about $\unit[600]{ps}$ prior to the THz pulse for exciton formation and cooling~\cite{ChatterjeePRL2004,ZybellAPL2011}, practically a pure occupation of the 1$s$ excitonic state is
prepared. 

Figure~\ref{Fig: intro}(b) shows the effect of THz and magnetic fields on the measured differential PL spectrum denoted as 
$\Delta \mathrm{PL} = \mathrm{PL}_{\mathrm{on}} - \mathrm{PL}_{\mathrm{off}}$ that is the difference of the PL with THz ($\mathrm{PL}_{\mathrm{on}}$) and without THz ($\mathrm{PL}_{\mathrm{off}}$) field. We have normalized the PL such that the THz-off case produces unity at the 1$s$-emission resonance.
The THz energy is set to $\unit[8.7]{meV}$ and is resonant with the 1$s$-to-2$p$ transition energy without magnetic field.
The black (red) line shows $\Delta \mathrm{PL}$ for $B=\unit[0]{T}$ ($B=\unit[2.1]{T}$), shortly ($\unit[16]{ps}$) after the THz-pulse center coincides with the QW. The spectrum above $\unit[5]{meV}$ (shaded area) has been rescaled by a factor of 30 to enhance the visibility of the 2$s$ peak.
We observe that $\mathrm{PL}_{1s}$ is quenched without $B$-field to a level that decreases only weakly as $B$ is increased to $\unit[2.1]{T}$.
Without $B$-field, we notice a well-pronounced 2$s$ enhancement at $\unit[0]{T}$, i.e., $\Delta \mathrm{PL}_{2s}>0$, which vanishes completely for $B=\unit[2.1]{T}$. To analyze 1$s$- and 2$s$-PL effects directly, we spectrally integrate $\Delta \mathrm{PL}$ over the regions indicated by the dashed vertical lines in Fig.~\ref{Fig: intro}(b).
This procedure isolates the THz-induced changes in the 1$s$ ($\Delta \mathrm{PL}_{1s}$) and 2$s$ ($\Delta \mathrm{PL}_{2s}$) emission.

\begin{figure}
\centering
\includegraphics[width=0.8 \columnwidth]{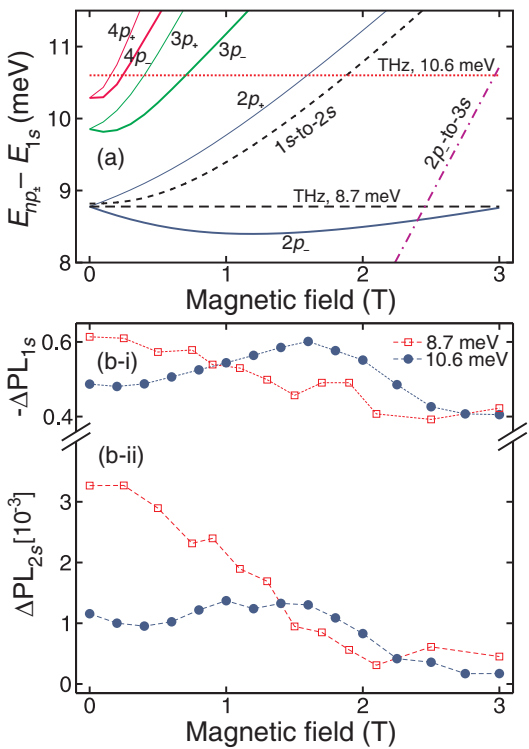}
\caption{(color online).
Magnetic-field control of intra-exciton transitions.
(a) Computed transition energy from 1$s$ to $np_{\pm}$ exciton states as a function of $B$. The dashed (dotted) horizontal line marks the THz central energy at $8.7$ ($10.6$) $\mathrm{meV}$. The dashed (dashed-dotted) curve presents the 1$s$-to-2$s$ (2$p_-$-to-3$s$) energy difference while $p_-$ ($p_+$) branches are shown as thick solid (thin solid) lines.
(b) Experimental data of 1$s$ quench (b-i) and 2$s$ excess PL (b-ii) as a function of $B$.
The THz energy is $\unit[8.7]{meV}$ for the squares and $\unit[10.6]{meV}$ for the circles; dashed lines are guide to the eye. 
}
\label{Fig: 1s-np transitions and experiment}
\end{figure}

Figure~\ref{Fig: 1s-np transitions and experiment}(a) shows the calculated dipole-allowed transition energies between 1$s$ and $p$-like states as a function of $B$-field, 
constructing a fan chart of magnetoexcitons~\cite{Maan1987,SCHMITTRINK1991}.
The $p$-like states are classified by the main quantum number $n\geq 2$, angular momentum quantum number $l=1$, and magnetic quantum number $m=-1$ ($m=+1$) for $np_-$ ($np_+$) states.
The 1$s$-to-$np_-$ (1$s$-to-$np_+$) exciton transition energies are shown as thick solid (thin solid) lines.
The 1$s$-to-2$s$ (2$p_-$-to-3$s$) energy difference is also presented as a dashed (dashed-dotted) curve.
We expect that scattering from 2$p_-$-to-2$s$ is weaker compared to 2$p_+$-to-2$s$
because the 2$s$ state (dashed curve) is energetically closer to the 2$p_+$ state.
In agreement with Refs.~\cite{Nickel2000,Barticevic2002,Mi2009}, the $p_-$ branch first red shifts before it is eventually blue shifted, while the $2p_+$ branch always shows a monotonically increasing blue shift~\cite{Nickel2000,Barticevic2002,Mi2009}.

In our experiments, we change the magnetic-field strength $B$ for a fixed THz energy $\hbar \omega_{\mathrm{THz}}=8.7$ (dashed horizontal line) or $\unit[10.6]{meV}$ (dotted horizontal line) and record $\Delta \mathrm{PL}_\lambda$ at the $\lambda=1s$ and $\lambda=2s$ resonances.
The $\hbar \omega_{\mathrm{THz}}=\unit[8.7]{meV}$ excitation energy is resonant with the 1$s$-to-2$p$ transition at $\unit[0]{T}$ while $\hbar \omega_{\mathrm{THz}}=\unit[10.6]{meV}$ is resonant with the 1$s$-to-2$p_+$ transition at $\unit[1.6]{T}$.
The $\hbar \omega_{\mathrm{THz}}=\unit[8.7]{meV}$ excitation energy is mostly resonant with the 2$p_-$ branch for elevated $B$ because the 2$p_+$ branch quickly becomes non-resonant for increased $B$. 
For $\hbar \omega_{\mathrm{THz}}=\unit[10.6]{meV}$, the 2$p_+$ state becomes resonant with THz transitions at $B=\unit[1.6]{T}$.

Figure~\ref{Fig: 1s-np transitions and experiment}(b-i) shows the measured maximum $-\Delta \mathrm{PL}_{1s}$ as a function of $B$
for $\hbar \omega_{\mathrm{THz}}=\unit[8.7]{meV}$ (squares) and $\hbar \omega_{\mathrm{THz}}=\unit[10.6]{meV}$ (circles).
The corresponding $\Delta \mathrm{PL}_{2s}(B)$ data are shown in Fig.~\ref{Fig: 1s-np transitions and experiment}(b-ii).
We have used a low-pass filter to remove the experimental noise within the time dynamics of $\mathrm{PL}(t)$.
For $\hbar \omega_{\mathrm{THz}}=\unit[8.7]{meV}$ and $B=\unit[0]{T}$, the THz field transfers 1$s$ excitons identically to both $2p_+$ and 2$p_-$ excitons because they are degenerate. 
This generates a large quench observed in $\Delta \mathrm{PL}_{1s}$.
For higher $B$-values, the 2$p_+$ state becomes non-resonant very fast such that only 2$p_-$ is near resonant with the THz field.
Hence, the 1$s$ quench results from the THz coupling between 1$s$ and 2$p_-$ states for elevated $B$, while the 2$p_+$ becomes uncoupled. 
Consequently, $\Delta \mathrm{PL}_{1s}$ drops monotonically by roughly 40\% as $B$ is elevated.
For the same conditions, $\Delta \mathrm{PL}_{2s}$ drops drastically by a factor of seven.
In particular, the large changes in $\Delta \mathrm{PL}_{2s}$ compared to moderate changes in $\Delta \mathrm{PL}_{1s}$ demonstrate directly that the $B$-field efficiently controls the Coulomb scattering that directly influences only the $\Delta \mathrm{PL}_{2s}$ part.

In addition, a new feature emerges to $\Delta \mathrm{PL}_{2s}$ starting at $B=\unit[2]{T}$ for $\hbar \omega_{\mathrm{THz}}=\unit[8.7]{meV}$: The monotonically decaying trend does not continue, but data points indicate the presence of a new resonance. On the basis of our microscopic 
theoretical analysis explained below, we assign this resonance to a resonant two-photon transition from 1$s$ to 3$s$ via the 2$p_-$ state and subsequent relaxation toward 2$s$.
Similar internal two-photon transitions have been observed in Ref.~\cite{KonoPRL1997} between 1$s$, 2$p$, and 2$s$ states for high $B$-fields up to \unit[12]{T}.
To distinguish the new 2$p_-$-to-3$s$ resonance from the scattering-induced 2$p$-to-2$s$ transfer, one must have a significantly weakened scattering $\Delta \mathrm{PL}_{2s}$, which is the case above $B=\unit[2]{T}$.

For the initially ($\unit[0]{T}$) detuned $\hbar \omega_{\mathrm{THz}}=\unit[10.6]{meV}$ excitation (circles), we observe a resonance both in $-\Delta \mathrm{PL}_{1s}$ and $\Delta \mathrm{PL}_{2s}$ at around $\unit[1.6]{T}$.
Figure~\ref{Fig: 1s-np transitions and experiment}(a) shows that 1$s$-to-2$p_+$ transition becomes then resonant while neither 2$p_+$ or 2$p_-$ become resonant as $B$ is detuned away from $B=\unit[1.6]{T}$.
This explains that the detuned case produces a $\Delta \mathrm{PL}_{1s}$ quenching resonance around $B=\unit[1.6]{T}$.
However, the quench behavior for $\hbar \omega_{\mathrm{THz}}=\unit[10.6]{meV}$ is asymmetric, yielding more quenching (50\%) at $B=\unit[0]{T}$ compared to $B=\unit[3]{T}$ (40\%). This is consistent with Fig.~\ref{Fig: 1s-np transitions and experiment}(a) because there are more nearby final states at $\unit[0]{T}$ than at $\unit[3]{T}$.
Also $\Delta \mathrm{PL}_{2s}$ shows a maximum at $\unit[1.6]{T}$, which follows as the 2$p_+$ population is transferred into 2$s$ population via the Coulomb scattering. 
As for the $\hbar\omega_{\mathrm{THz}}=\unit[8.7]{meV}$ excitation, the Coulomb scattering is reduced for elevated $B$
such that $\Delta \mathrm{PL}_{2s}(B)$ decreases for $B>\unit[1.6]{T}$.
At the same time, the maximum of $\Delta \mathrm{PL}_{2s}(B)$ remains smaller compared with $\hbar \omega_\mathrm{THz}=\unit[8.7]{meV}$ case because the Coulomb scattering is already significantly reduced at the peaking $B=\unit[1.6]{T}$ field.

For our microscopic analysis, we start from the standard many-body Hamiltonian that includes the electronic band structure, the Coulomb interactions among the charge carriers, as well as the light-field and THz interactions~\cite{PQE2006,SQObook2011}.
To account for the $B$-field, we use the Hamiltonian~\cite{Boettge2013}
\begin{align*}
\notag
\hat H_B &= \frac{\mathbf{\hat p}^2}{2\mu} + \frac{\mathbf{\hat P}^2}{2M} - V(\mathbf{r}) + \frac{\mu}{2} \boldsymbol{\bar \omega}_\mu^2 \mathbf{r}_{\parallel}^2 
\\
&+ \frac{|\boldsymbol{\bar \omega}_\mathrm{e}| - |\boldsymbol{\bar \omega}_\mathrm{h}|}{2} \hat L_z + \boldsymbol{\bar \omega}_M \cdot (\mathbf{r} \times \mathbf{\hat P})
\end{align*}
which leads to the magnetoexcitons~\cite{SCHMITTRINK1991} when solving the generalized Wannier equation.
Here, the relative (center-of-mass) coordinate is denoted as $\mathbf{r}$ ($\mathbf{R}$) with momentum 
$\mathbf{\hat p}$ ($\mathbf{\hat P}$) and QW in-plane component $\mathbf{r}_{\parallel}$. The reduced mass ($\mu$) and total mass ($M$) enter together with the Coulomb interaction $V(\mathbf{r})$, effective cyclotron frequencies $\boldsymbol{\bar \omega}_j$, where $\mathrm{e}$ and $\mathrm{h}$ denote electron and hole, respectively, and angular momentum operator $\hat L_z$.
We solve the exciton dynamics~\cite{Boettge2013,Rice2013} in the presence of THz and $B$-fields for all relevant bright and dark exciton states and compute the resulting PL via the Elliott formula.

\begin{figure}
\centering
\includegraphics[width=0.78 \columnwidth]{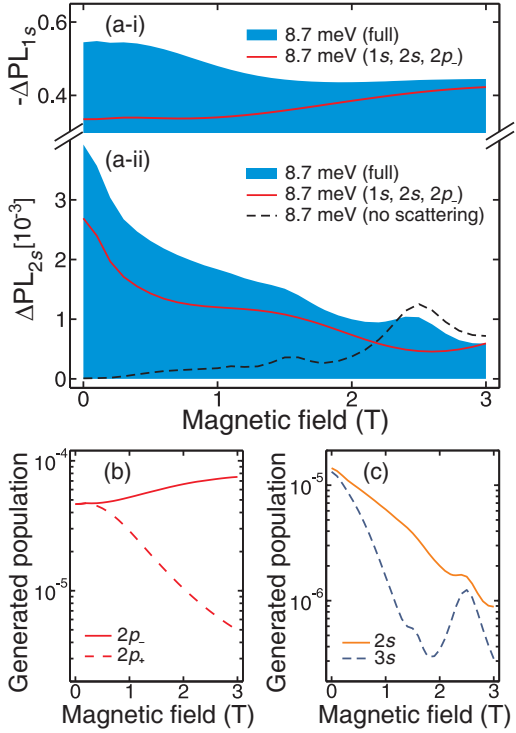}
\caption{(color online).
Computed magnetic-field control of intra-exciton transitions.
Computed 1$s$ quench (a-i) and 2$s$ excess PL (a-ii) as a function of $B$. The full (shaded) calculation is compared with separate calculations based on a reduced THz-current matrix (dark solid) and without scattering (dashed).
(b)-(c) Generated $p$-like (b) and $s$-like (c) exciton populations as a function of $B$, based on the full calculation. 
In all frames, the THz central energy is $\unit[8.7]{meV}$ and resonant with the 1$s$-to-2$p$ transition at $\unit[0]{T}$.
}
\label{Fig: switch on-off 8.7 meV and populations}
\end{figure}

Figure~\ref{Fig: switch on-off 8.7 meV and populations} shows the computed $-\Delta \mathrm{PL}_{1s}$ [Fig.~\ref{Fig: switch on-off 8.7 meV and populations}(a-i)] and $\Delta \mathrm{PL}_{2s}$ [Fig.~\ref{Fig: switch on-off 8.7 meV and populations}(a-ii)] as a function of $B$ for the $\hbar \omega_{\mathrm{THz}}= \unit[8.7]{meV}$ excitation.
The full calculation (shaded area) is compared with calculations having a reduced amount of THz transitions (dark solid) and vanishing Coulomb scattering (dashed line).
More specifically, we reduce the number of THz transitions by including only transitions between the states 1$s$, 2$s$, and 2$p_-$. 
The results of the full calculations (shaded area) agree well with the $\hbar \omega_{\mathrm{THz}} = \unit[8.7]{meV}$ excitation 
[Fig.~\ref{Fig: 1s-np transitions and experiment}(b) vs. Fig.~\ref{Fig: switch on-off 8.7 meV and populations}(a)].

On the basis of this quantitative agreement, we can proceed to identify the effect of the individual interaction processes on the measured $B$-field control.
For example, excluding the transitions to 2$p_+$ (Fig.~\ref{Fig: switch on-off 8.7 meV and populations}(a-i), dark line) reduces  the 1$s$ quench by approximately a factor of $\sqrt{2}$ at $B=\unit[0]{T}$ because only \textit{one} 2$p$ state is available for this particular calculation. For high $B$, however, the $p_+$-excluded calculation approaches the full calculation (shaded area).
However, the exclusion of 2$p_+$ and 3$s$ does not change the $\Delta \mathrm{PL}_{2s}$ increase much, as seen from Fig.~\ref{Fig: switch on-off 8.7 meV and populations}(a-ii).
As the major difference, only the full computation (shaded area) shows a peak around $B=\unit[2.5]{T}$.
The resonance is recovered only if 3$s$ transitions are included making the resonant 2$p_-$-to-3$s$ transition possible,
see also the 2$p_-$-to-3$s$ energy difference [dashed-dotted curve in Fig.~\ref{Fig: 1s-np transitions and experiment}(a)] that becomes resonant with the THz field at $B=\unit[2.5]{T}$.
When the scattering is switched off while the 1$s$-2$p_-$-3$s$ transition is fully included [Fig.~\ref{Fig: switch on-off 8.7 meV and populations}(a-ii), dashed line], only the resonance close to $\unit[2.5]{T}$ remains, which shows that it is not related to scattering but to a 1$s$-2$p_-$-3$s$ transition.
These switch-off analyses confirm conclusively that the experimental $\Delta \mathrm{PL}_{2s}$ peak in Fig.~\ref{Fig: 1s-np transitions and experiment}(b-ii) indeed originates from the eventual transition to the 3$s$ state.

We have also computed the peak differences in the THz-generated exciton population $\Delta N_\lambda$.
Figure~\ref{Fig: switch on-off 8.7 meV and populations}(b) analyzes $\Delta N_{2p_\pm}$ and Fig.~\ref{Fig: switch on-off 8.7 meV and populations}(c) shows $\Delta N_{2s}$ and $\Delta N_{3s}$ as function of $B$.
We observe that the THz field dominantly excites a 2$p_-$ population that increases slightly for elevated $B$.
We also see that the 2$p_\pm$ populations are the same only at $\unit[0]{T}$.
At the same time, the 2$s$ population decreases by more than one order of magnitude because the increased $B$-field decreases the Coulomb scattering.
The THz-induced $\Delta N_{3s}$ displays a clear resonance at $\unit[2.5]{T}$, providing independent evidence that the experimental $\Delta \mathrm{PL}_{2s}$ resonance stems from the 1$s$-2$p_-$-3$s$ transition.

\begin{figure}
\centering
\includegraphics[width=0.78 \columnwidth]{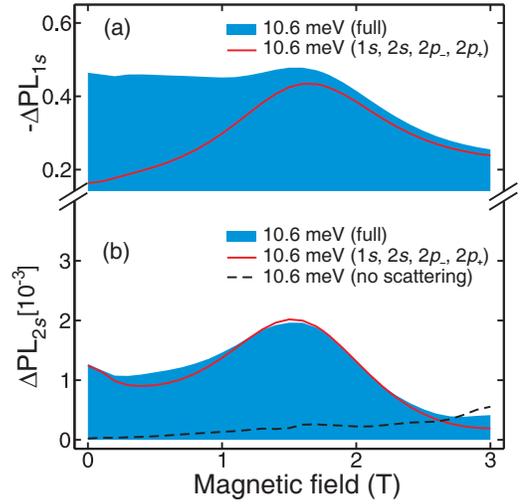}
\caption{(color online).
Computed magnetic-field control for $\hbar \omega_{\mathrm{THz}} = \unit[10.6]{meV}$.
Computed 1$s$ quench (a) and 2$s$ excess PL (b) as a function of $B$. The full (shaded) calculation is compared with separate calculations based on a reduced THz-current matrix (dark solid) and vanishing scattering (dashed).
}
\label{Fig: switch on-off 10.6 meV}
\end{figure}

Figure~\ref{Fig: switch on-off 10.6 meV} presents a similar analysis as in Fig.~\ref{Fig: switch on-off 8.7 meV and populations}(a), but now for the initially detuned THz excitation $\hbar \omega_{\mathrm{THz}} = \unit[10.6]{meV}$.
The computed $-\Delta \mathrm{PL}_{1s}$ [(a)] and $\Delta \mathrm{PL}_{2s}$ [(b)] are shown based on the full (shaded area) and reduced  calculations with a four-level model containing only 1$s$, 2$s$, and 2$p_\pm$ states (dark solid line) and without scattering (dashed line).
Also for this detuning, the full computation reproduces the experimental $-\Delta \mathrm{PL}_{1s}$ and $\Delta \mathrm{PL}_{2s}$ presented in Fig.~\ref{Fig: 1s-np transitions and experiment}(b) (circles).

The switch-off analysis yields further insight to the relevant processes. For example, the four-level calculation (dark solid line) yields large differences for $-\Delta \mathrm{PL}_{1s}$ for $B$ below $\unit[1]{T}$. This follows because the higher exciton branches are near-resonant with the THz field for $\hbar \omega_{\mathrm{THz}}=\unit[10.6]{meV}$ for low-enough $B$, see Fig.~\ref{Fig: 1s-np transitions and experiment}(a). 
At the same time, $\Delta \mathrm{PL}_{2s}$ is modified only slightly because it originates mainly from the Coulomb scattering between 2$p_\pm$ and 2$s$ already included in the four-level analysis.
We conclude that the dipole-allowed THz-induced transitions 1$s$-to-3$p_-$, 1$s$-to-3$p_+$, and to higher $p$-like states lead to enhanced quenching at $B=\unit[0]{T}$ compared to $B=\unit[3]{T}$. We also recognize that the theory curve [Fig.~\ref{Fig: switch on-off 10.6 meV}(a), shaded] reveals a similar asymmetric behavior as the experiment [Fig.~\ref{Fig: 1s-np transitions and experiment}(b-i), circles].
For $B$-fields beyond $\unit[1.6]{T}$, the full and four-level calculations are similar because only the four included states remain near resonant such that the $B$-field renders the system four-level like.
By omitting the Coulomb scattering, the computation almost completely suppresses $\Delta \mathrm{PL}_{2s}$ (dashed line) similar to Fig.~\ref{Fig: switch on-off 8.7 meV and populations}(a-ii) while $-\Delta \mathrm{PL}_{1s}$ is almost unchanged (not shown).
Hence, the diffusive Coulomb scattering is essential for all excitation conditions to correctly describe the scattering-induced 2$p$-to-2$s$ population transfer.

In summary, our time-resolved terahertz quenching studies of the excitonic photoluminescence show that the intra-excitonic transitions and interactions can be efficiently controlled via an external magnetic field.
Three major effects have been observed: The magnetic field modifies the Coulomb scattering, induces resonant two-photon THz transitions, and renders the system few-level-like for high field strengths. 
These observations lead to new possibilities to control THz transitions and Coulomb-scattering effects.

The Dresden group thanks P. Michel, W. Seidel and the FELBE team for their dedicated support. The work in Dresden (JB) was partially supported by the DFG project SCHN 1127/2-1.
The Marburg group thanks for financial support by the DFG project KI 917/2-1
and RTG 1782.



%

\end{document}